\documentclass[12pt,preprint]{aastex}

\slugcomment{}

\shorttitle{The First Distance to an Extrasolar Planet}  

\begin{document}


\title{The First Direct Distance and Luminosity 
Determination for a Self-Luminous
Giant Exoplanet: The Trigonometric Parallax to 2MASSW J1207334-393254Ab}


\author{B.A. Biller$^1$ $\&$ L.M. Close$^1$}

\email{bbiller@as.arizona.edu}

\affil{$^1$Steward Observatory, University of Arizona, Tucson, AZ 85721}



\begin{abstract} 
We present the first trigonometric parallax and distance for a young
planetary mass object.  A likely TW Hya cluster member, 
2MASSW J1207334-393254Ab (hereafter 2M1207Ab) is
an M8 brown dwarf with a mid to late L type 
planetary mass companion.  Recent observations of spectral
variability have uncovered clear signs
of disk accretion and outflow, 
constraining the age of the system to $<$10 Myr. Because
of its late spectral type and the clearly youthful nature of the
system, 2M1207b is very likely a planetary mass object.   
We have measured the first accurate distance and
luminosity for a self-luminous planetary mass object.  
Our parallax measurements are accurate to $<$2 mas (1$\sigma$) 
for 2M1207Ab.
With 11 total epochs of data taken from January 2006 through April 2
007 (475 images for 2M1207Ab), we determine
a distance of 58.8$\pm$7.0 pc (17.0$^{+2.3}_{-1.8}$ mas, 1.28$\sigma$) 
to 2M1207Ab and a calculated luminosity of 
0.68-2.2$\times$10$^{-5}$ L$_{\odot}$ for 2M1207b. 
Hence 2M1207Ab is a clear member of the TW Hya cluster in terms of 
its distance, proper motions, and youthful nature.  However, as 
previously noted by Mohanty and co-workers, 2M1207b's luminosity 
appears low compared to its temperature according to evolutionary models.
\end{abstract}


\section {Introduction}
A very likely member of the $\sim$8 Myr TW
Hydrae association (at distances of 35-70 pc, Mohanty et al. 2003, Mamajek 
2005), 2MASSW J1207334-393254A (hereafter 2M1207A)
is a young M8 brown dwarf with a late L companion (2M1207b, Chauvin et
al. 2004, spectral type from Mohanty et al. 2007). Recent
observations of spectral variability (Scholz and Jayawardhana 2006)
have revealed accretion and jet-like 
features, constraining the age of the system
to $<$10 Myr; with such a young age and low temperature, 
2M1207b is very likely a planetary mass object.

Interestingly, 
previous distance estimates for this object have produced somewhat conflicting
results. Using the measured K band magnitudes and extrapolating a K
band absolute magnitude for a 8 Myr brown dwarf (from trends in Song
et al.  2003), Chauvin et al. (2004) estimated a photometric
distance of $\sim$70 pc at the outer edge of TW Hya from their
extrapolated distance modulus. At this age and distance, the models of
Baraffe et al. (2001) predict that the M8 dwarf should have a mass of
25 M$_{Jup}$ and the companion should have a mass of $\sim$5
M$_{Jup}$. Mamajek (2005) estimates a theoretical moving cluster distance to
2M1207 of 53$\pm$6 pc and infers masses for 2M1207A and b (with an
age of 8 Myr) of $\sim$21 M$_{Jup}$ and $\sim$3-4 M$_{Jup}$
respectively.  From precision HST proper motions, 
Song et al. (2006) 
also estimate a similar moving cluster 
distance -- 59$\pm$7 pc, and hence similar masses. 
However, using these closer distance estimates, 2M1207A becomes
underluminous and falls nearer the locus of the (120 Myr) Pleiades on a
color-magnitude diagram (Fig. 1) -- which is inconsistent with an
object age of $<$10 Myr at $\sim$50 pc.
Recently, Mamajek and Meyer (2007) have revised the estimated
theoretical moving cluster distance to 2M1207 to 66$\pm$5 pc. 
Thus, a direct distance measurement via parallax would help clarify this
situation and additionally would constrain a number of
important properties for this object.  Since the youth ($<$10 Myr) 
and hence low mass nature (M$<$13 M$_{Jup}$) of 2M1207b has recently been
confirmed, we have also measured the first accurate luminosity for a
self-luminous planetary mass object.

\section{Observations and Data Reduction}

We have acquired 11 epochs of data stretching from January 2006
to April 2007 with ANDICAM at the SMARTS 1.3 m telescope on Cerro Tololo.
Both 2M1207 and the standard object LHS 2397a (an M8 dwarf with 
an L7.5 brown dwarf companion with a well known parallax of
62.3$\pm$4.0 mas -- Tinney (1996)) were observed for 
40 minutes over transit.  Observations 
were repeated over multiple nights bracketing the new moon each month.
Over the entire observing period, 475 I band data frames with per frame
exposure time of 300 s were acquired for 2M
1207 and 491 I band data frames were acquired for LHS 2397a.

The M8 primary of 2M1207
has I=15.8 (Scholz et al. 2005) so
parallax can be determined in the visible.  The
L5-9.5 secondary of 2M1207 has K=16.9 (Chauvin et al. 2004), is
even fainter in the visible (I$>$19), and lies within 0.8" of its
primary, thus it is essentially invisible to the 1.3m in the optical (and
is not apparent in our images). Its presence does not affect our
attainable astrometric precision. LHS 2397a is comparably bright as
2M1207 and was chosen as a standard object due to the fact that
both are late M dwarf + mid to late L dwarf binaries (Freed et al. 2003).   
Our measurement of the parallax of LHS 2397a (previously found to be 
62.3$\pm$4.0 mas, Tinney (1996)) serves as  
a test of our observational and data reduction procedures.
Each target object was always placed at 
82 pix E and 148 pix N of the center of the chip at pixel 
(330, 660).  A number of bright stars lie right outside the LHS 2397a field;
we chose to place our target objects in the upper left quadrant of the chip
as opposed to the center of the chip in order to keep these bright stars
and their saturation bleeds off of the CCD.

A dedicated parallax data analysis pipeline was used to reduce these data.
This pipeline aligns each data frame to a master frame, removes cosmic
rays, and performs PSF fitting
photometry for 139 stars per frame using the DAOphot allstar task 
(Stetson 1987).  The ANDICAM optical detector is a Fairchild 447 
2048$\times$2048 CCD and was used in 2$\times$2 binning mode
yielding a nominal 
platescale of 0.369 arcsec/pixel (ANDICAM website).  Five bright stars 
in different parts of the field 
were used to calibrate x and y plate scale changes on the CCD.  The
separations between the 5 platescale calibration stars were measured for 
each data frame and then normalized to the average value over all frames.
These normalized values were used as frame by frame platescale corrections.
We found platescale variations of 
less than 0.06$\%$ over the $\sim$1.5 year time baseline of our observations''

The position of 2M1207 and LHS 2397a in RA and DEC as a function of 
time is shown in Figs.~\ref{fig:parallax}. 
Many reference stars as bright or brighter than the target 
are available
in both the 2M1207 and LHS 2397a fields; thus, we have a good
distribution of baselines around both objects from which to calculate 
parallax.  The position of 2M1207 and LHS
2397a were measured relative to 40 and 15 reference stars
respectively.

To correct for a slightly variable x and y  
``pincushion'' distortion (first derivative of the x and 
y platescale) on the chip, we measured the apparent motion 
on the sky for the 5 stars nearest 2M1207 on the chip.
These stars are faint and likely significantly behind 2M1207 
and thus presumed background so the motion observed for these
should calibrate the local distortion of the chip near our target.  
Two of these stars showed 
apparent space motions of their own and were thus discarded.
The motions from the remaining three stars were then averaged together
to create a first derivative platescale correction curve.  
This correction curve was then 
subtracted from the measured parallax curve for 2M1207.  A similar 
correction was also performed for the LHS 2397a data.

To determine the trigonometric 
parallax to 2M1207 and LHS 2397a, we fit (least 
squares fit) our measured parallax data to a parallax and 
proper motion model for each dataset.
Precise HST proper motions for 2M1207 
of $\mu_{\alpha}$ = -60.2$\pm$4.9 mas/yr and $\mu_{\delta}$ = 
-25.0$\pm$4.9 mas/yr were adopted from Song et al. (2006).
Proper motions for LHS 2397a of $\mu_{\alpha}$ = -508.0$\pm$20.0 
mas/yr and $\mu_{\delta}$ = -80.0$\pm$20.0 mas/yr 
were adopted from Salim $\&$ Gould (2003).
Fits were performed both holding proper motion fixed and also fitting 
to proper motion; for both 2M1207 and LHS 2397a we retrieve the 
published proper motions to within the published errors.
We find that the error in published proper motion is negligible
compared to per frame measurement errors.
Error in each nightly parallax measurement was calculated from the rms 
of measurements taken over that night.  
Measurements from late 2006 
(2006.9 -- 2007.1 epoch) were discarded since these 
data were taken off transit and 
thus suffer considerably from Differential Color Refraction 
(DCR, see, e.g. Dahn et al. 2002).

In order to estimate the error in our trigonometric 
parallax measurement, a Monte-Carlo 
ensemble of 10000 datasets was simulated by multiplying 
 our measured position errors by a random number distribution pulled from 
a Gaussian distribution, then adding that random error to the measured 
position.  An error-free trigonometric parallax model was then fitted and 
chi-square minimized to each of these simulated observations.  The adopted 
distance is the mean of this 10000 simulated observation distribution of fits 
and the adopted error is the 1.28$\times$ the 
standard deviation of this distribution, corresponding to an 80$\%$
confidence interval.  
Histograms of the distributions of observation fits are presented in 
Fig.~\ref{fig:hist}. 

We estimate a correction from relative 
to absolute parallax of 1.2$\pm$0.9 mas, based on 
I band photometry of 40 reference stars for 2M1207 and 15 reference 
stars for LHS 2397a.  This correction was obtained by 
adopting an average spectral type of M0 for our relatively faint field 
reference stars (I = 15 - 18 mag), calculating photometric parallaxes 
for each reference star and then 
employing the median photometric parallax of the reference stars
as the estimated correction.

\section{Results and Discussion}

For 2M1207Ab, we acquired a best fit relative parallax of 
15.8$^{+2.1}_{-1.6}$ mas, corresponding to an 
absolute parallax of 17.0$^{+2.3}_{-1.8}$ mas and a 
best fit distance of 58.8$\pm$7.0 pc (all 1.28$\sigma$ errors).  For our
standard LHS 2397a, we acquired a best fit relative parallax of 
66.7$^{+5.2}_{-4.6}$ mas, corresponding to an absolute parallax of 
67.9$^{+5.3}_{-4.7}$ mas (similar to the previous result of 62.6$\pm$4.0 mas, 
Tinney 1996) and a best fit distance of 14.7$\pm$1.0 pc.
Fewer reference stars were available for the LHS 2397a 
standard object than for 2M1207, leading to a lower precision result.
Parallax results are presented in Table 1.  Absolute magnitudes derived 
using our measured distance are presented in Table 2.

Adopting an apparent J magnitude of 13.00$\pm$0.03 
(Mohanty et al. 2007) and BC$_J$=2.0 for a M8 (Dahn et al. 2002), 
we estimate a total
luminosity for 2M1207A of 2.7$\times$10$^{-3}$ L$_{\odot}$. 
Adopting an apparent J magnitude of 20.00$\pm$0.02 
(Mohanty et al. 2007) and BC$_J$=1.5 for a late L dwarf  
(Dahn et al. 2002), we estimate a total luminosity for 2M1207b 
of 6.8$\times$10$^{-6}$ L$_{\odot}$. 
We repeated this calculation in the Ks band: adopting m$_{Ks}$=11.95$\pm$0.03 
(Chauvin et al. 2004)
and BC$_K$=3.2 (Golimowski et al. 2004)
for 2M1207A and converting m$_{Ks}$ to m$_K$ using the 
transformations from Carpenter (2001), we estimate a total luminosity of 
2.4$\times$10$^{-3}$ L$_{\odot}$, consistent with the J band estimate.  
However, adopting m$_{Ks}$=16.93$\pm$0.11 (Chauvin et al. 2004) 
and BC$_K$=3.3-3.4 (Golimowski et al. 2004) for 2M1207b and converting
m$_{Ks}$ to m$_K$ using the transformations from Stephens \& Leggett (2004), 
we estimate 
a total luminosity of 2.0-2.2$\times$10$^{-5}$ L$_{\odot}$ -- 3$\times$ 
brighter than 
the J band estimate.  The culprit here is likely the bolometric corrections 
used -- which, while appropriate for older field objects are not entirely 
appropriate for this very young, very cool object.  
Accordingly, we do not assign error 
bars to our luminosity estimates because of the uncertainties in the 
bolometric corrections.  2M1207b is somewhat redder (J-K)
than field objects of the same spectral types.  Thus, BC$_K$ is especially
suspect, since the K band flux of 2M1207b represents a larger portion of 
its total bolometric flux than is true for comparable spectral type field 
objects.  Estimated luminosities are 
presented in Table 2.

We estimated mass and effective temperatures for 2M1207Ab using only 
our derived absolute magnitudes and the 
DUSTY models of Chabrier et al. (2000) and Baraffe et al. (2001).  
Adopting an isochronal age for the 
TWA Hydra cluster of $\sim$8$^{+4}_{-3}$ Myr (Song et al. 2003; 
Zuckerman \& Song 2004; Chauvin et al. 2004), we compared the absolute
J, H, K$_s$, 
and L$^{\prime}$ colors (after converting from 2MASS to CIT magnitudes) 
to the 5 and 10 Myr isochrones.  2M1207A 
is consistent in color and luminosity with the 20-30 M$_{Jup}$ models.  
2M1207b is roughly consistent in luminosity 
with the 3-7 M$_{Jup}$ models (early T spectral types) --
however, it possesses much redder colors than these models, 
consistent with mid-to-late L dwarfs (6-10 M$_{Jup}$ models).  This 
color/luminosity mismatch has been previously noted by Mohanty et al. (2007)
among others -- 2M1207b possesses a mid-late L spectrum and colors, 
but is underluminous for its age, possessing the luminosity expected of 
an early T, yet no methane absorption is observed.  
From spectral fitting, T$_{eff}$$\sim$1600 K for 2M1207b 
(Mohanty et al. 2007), yet we derive
an incorrect model T$_{eff}$ of only 1260-1430 K.  Put in other words, 
according to the models 2M1207b is 10$\times$ too faint for its 
spectral type and age.  A comparison of model to observed properties is 
presented in Table 3.

At a distance of $\sim$50 pc, both 2M1207A and b are somewhat 
underluminous for their respective spectral types (see Fig.~1).  
Increasing the distance to 59 pc and adopting T$_{eff}$=2550$\pm$150 pc 
(Mohanty et al. 2007) and age=5-10 Myr solves the underluminosity issue for 
2M1207A, which becomes consistent with the 30 M$_{Jup}$, 10 Myr DUSTY 
models to within 0.2 mag.   
However, adopting T$_{eff}$=1600$\pm$100 K (Mohanty et al. 2007) and 
age=5-10 Myr, at a distance of 59 pc, 
2M1207b is still underluminous compared to the models by 2-3 mag in 
JHK$_{s}$L$^{\prime}$.  Thus, our increase in distance does not resolve
the issue of 2M1207b's underluminosity.    

A number of reasons for the lower than expected luminosity of 2M1207b 
have been suggested.
Mohanty et al. 2007 suggest that an edge-on disk around 2M1207b produces 
$\sim$2.5 mag of gray extinction (over JHK$_s$L$^{\prime}$) 
and hence, the observed low luminosity.
Mamajek and Meyer (2007) suggest that 2M1207b may be a hot protoplanet
collision remnant.  Additionally, while equally unlikely, 2M1207b may not 
be coeval with 2M1207A and may indeed be an older, smaller 
hence less luminous captured L dwarf (or rather, 2M1207A was captured
by an old $\sim$60 M$_{Jup}$ L dwarf).  Indeed, the 
measured colors, absolute magnitudes (JHK$_s$L$^{\prime}$), and luminosity
of 2M1207b 
are consistent with that of a 10 Gyr 67 M$_{Jup}$ object with L spectral type
and T$_{eff}$=1500.

The culprit could simply be the initial conditions of the evolutionary 
models used to derive physical properties.
Marley et al. (2007) have noted that the ``hot-start'' evolutionary models 
for planets and brown dwarfs such as those from both the Lyon and 
Tucson groups (Baraffe et al. 2003, Burrows et al. 2003) possess very high 
initial entropies and predict considerably
brighter luminosities for young high mass 
planets than models which start with lower
entropy initial conditions (which may be more appropriate for planets which 
form via core accretion).  For 4-10 M$_{Jup}$ objects, 
the Marley et al. (2007) models converge
with the standard evolutionary models by 100 Myr.
While 2M1207b most likely formed via fragmentation from a cloud core 
rather than core accretion or gravitational collapse within 2M1207A's 
small disk, the initial entropy conditions of its formation might 
have been considerably lower than those utilized by 
standard evolutionary models, 
producing a lower luminosity for each spectral type than expected at 
very young ages.  In particular, inside a binary system, the initial 
entropy conditions may have been different, presumably lower, 
for a forming 8 M$_{Jup}$ object
interacting with a 30 M$_{Jup}$ ``primary'' than for a 8 M$_{Jup}$
object forming individually.
Mohanty 
et al. (2007) claim that the models are not the culprit for 2M1207b's 
underluminous nature, comparing it with the young, low mass brown dwarf 
AB Pic B, whose colors, T$_{eff}$, and luminosity agree well with the models.
However, AB Pic is somewhat older than 2M1207Ab (30 Myr 
vs. $<$10 Myr) and has a much wider separation between components 
(250 AU vs. 50 AU), so it may have already converged to the 
standard evolutionary model tracks.

\section{Conclusions}

We measured a distance of 58.8$\pm$7.0 pc (17.0$^{+2.3}_{-1.8}$ mas) 
to 2M1207Ab and a luminosity of 
0.68-2.2$\times$10$^{-5}$ L$_{\odot}$ for 2M1207b, 
making 2M1207Ab a clear member of the TW Hya cluster.  While 
2M1207A now agrees well with evolutionary models, 2M1207b is still 
2-3 mag underluminous in JHK$_{s}$L$^{\prime}$ 
for its suggested T$_{eff}$ of 1600$\pm$100 K 
(Mohanty et al. 2007) and age of $\sim$8$^{+4}_{-3}$ Myr 
(for the TW Hya cluster, Song et al. 2003). 

\acknowledgements

This publication is based on observations made with the NOAO 1.3m telescope 
operated by the SMARTS consortium.  We acknowledge the excellent telescope 
queue support through the SMARTS consortium and would especially like to thank
Jenica Nelan, Charles Bailyn, Juan Espinoza, David Gonzalez, and 
Alberto Pasten.  We thank Matt Kenworthy for the suggestion of the 
Monte Carlo observation fits and Eric Mamajek for useful suggestions.  
BAB was supported by the NASA GSRP grant 
NNG04GN95H and NASA Origins grant NNG05GL71G. LMC is supported by an NSF 
CAREER award and the NASA Origins of the Solar System program.

\clearpage

\begin{deluxetable}{lccccc}
\tablecolumns{5}
\tablewidth{0pc}
\tablecaption{Parallax Results}
\tablehead{
\colhead{Target} & \colhead{RA} & \colhead{DEC} & \colhead{$\pi$(relative)} & 
\colhead{$\pi$(absolute)} & \colhead{Distance} }
\startdata 
2M1207Ab & 12 07 33.4 & -39 32 54.0 & 15.8$^{+2.1}_{-1.6}$ mas & 17.0$^{+2.3}_{-1.8}$ mas & 58.8$\pm$7.0 pc \\
LHS 2397a & 11 21 49.2 & -13 13 08.4 & 66.7$^{+4.8}_{-4.2}$ mas & 67.9$^{+5.3}_{-4.7}$ mas & 14.7$\pm$1.0 pc \\
\enddata
\end{deluxetable}

\begin{deluxetable}{lccccc}
\tablecolumns{6}
\tablewidth{0pc}
\tablecaption{Properties of 2M1207Ab}
\tablehead{
\colhead{Target} & \colhead{M$_J$\tablenotemark{a}} & \colhead{M$_H$\tablenotemark{b}} & \colhead{M$_{Ks}$\tablenotemark{b}}& \colhead{M$_{L}$\tablenotemark{b}} & \colhead{Luminosity (L$_{obs}$)}}
\startdata
2M1207A & 9.15$^{+0.28}_{-0.24}$ & 8.54$^{+0.28}_{-0.24}$ & 8.10$^{+0.28}_{-0.24}$ & 7.53$^{+0.29}_{-0.24}$ & 2.4-2.7$\times$10$^{-3}$ \\
2M1207b & 16.15$^{+0.34}_{-0.31}$ & 14.24$^{+0.35}_{-0.32}$ & 13.08$^{+0.30}_{-0.26}$ & 11.43$^{+0.31}_{-0.28}$ & 0.68-2.2$\times$10$^{-5}$ \\
\enddata
\tablenotetext{a}{Apparent magnitude from Mohanty et al. 2007}
\tablenotetext{b}{Apparent magnitude from Chauvin et al. 2004}
\end{deluxetable}

\begin{deluxetable}{lcccc}
\tablecolumns{5}
\tablewidth{0pc}
\tablecaption{Comparison of Measured to Model Properties (5-10 Myr DUSTY models)}
\tablehead{
\colhead{Target} & \colhead{T$_{effobs}$\tablenotemark{a}} & \colhead{T$_{effmodel}$\tablenotemark{b}} & \colhead{Mass from L$_{obs}$} & \colhead{Mass from T$_{effobs}$\tablenotemark{a}}}
\startdata
2M1207A & 2550$\pm$150 K & 2500-2700 K & 20-30 M$_{Jup}$ & 20-30 M$_{Jup}$ \\ 
2M1207b & 1600$\pm$100 K & 1260-1430 K & 3-7 M$_{Jup}$ & 6-10 M$_{Jup}$ \\
\enddata
\tablenotetext{a}{from Mohanty et al. 2007}
\tablenotetext{b}{T$_{eff}$ derived from age and luminosity (L$_{obs}$).}
\end{deluxetable}

\clearpage





\begin{figure}
 \includegraphics[angle=0,width=3in]{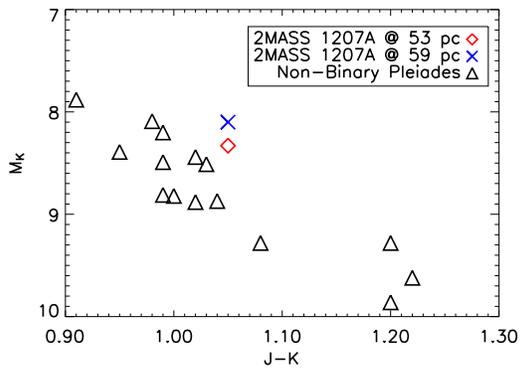}
\caption{M$_{K}$ vs. J-K HR diagram.  
Previous distance estimates for 2M1207A have led to a number of
inconsistencies.  J and K magnitudes are taken from the 2MASS survey.
Triangle points are non-binary Pleiades members
from Mart\'in et al. (2000).  At the Mamajek (2005) 
moving cluster distance of 53 pc, 2M1207A falls very close to the locus of
the Pleiades (120 Myr).  This is inconsistent with the age of $<10$
Myr for this object unless this object is extincted by circumstellar
dust.  At a distance of 59 pc, 2M1207A is considerably above the 
locus of the Pleiades, consistent with an age of $<10$ Myr.}
\label{fig:HR}
\end{figure}

\begin{figure}
\begin{center}
\begin{tabular}{cc}
\includegraphics[width=3.in]{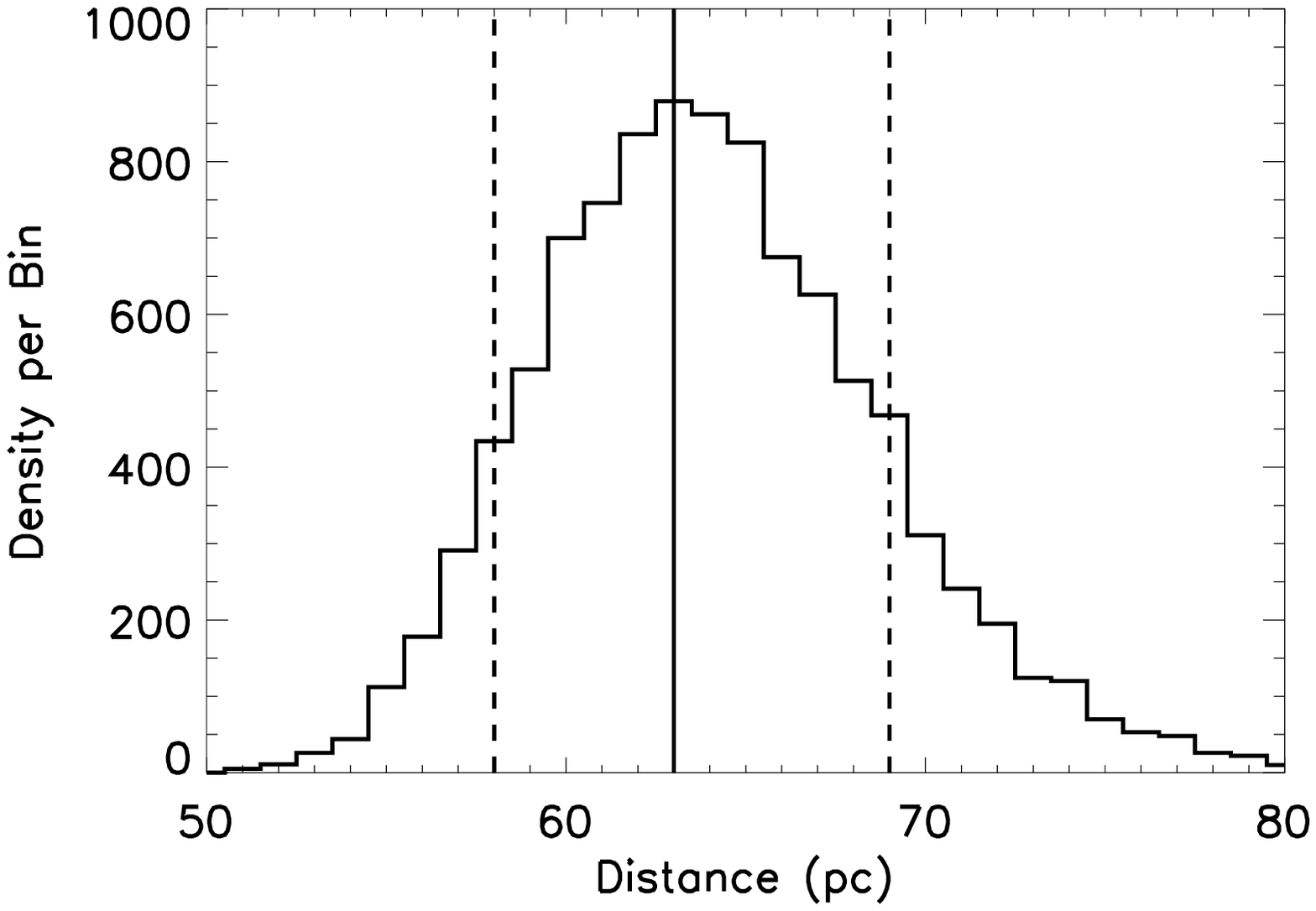} &
\includegraphics[width=3.in]{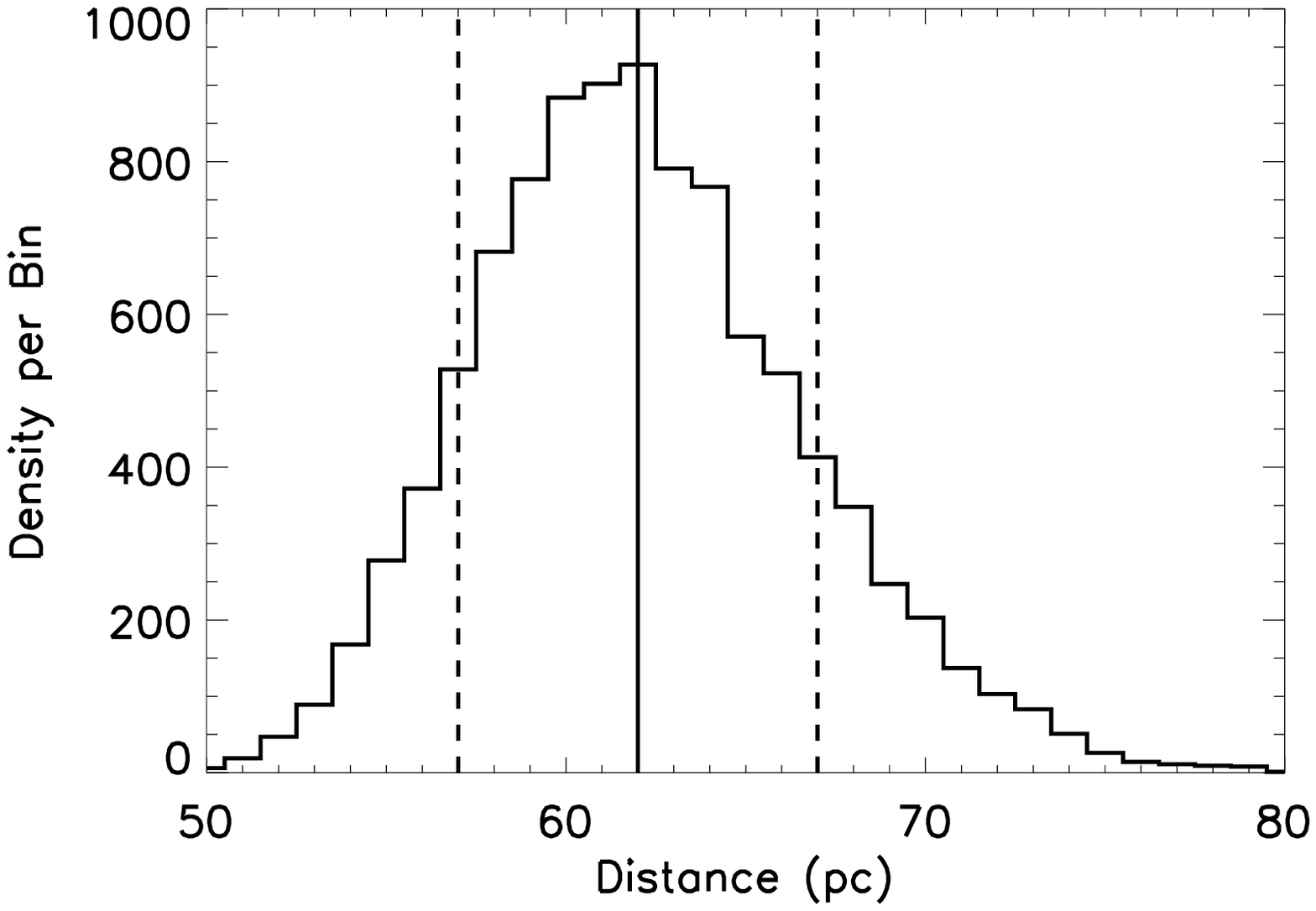} \\
\includegraphics[width=3.in]{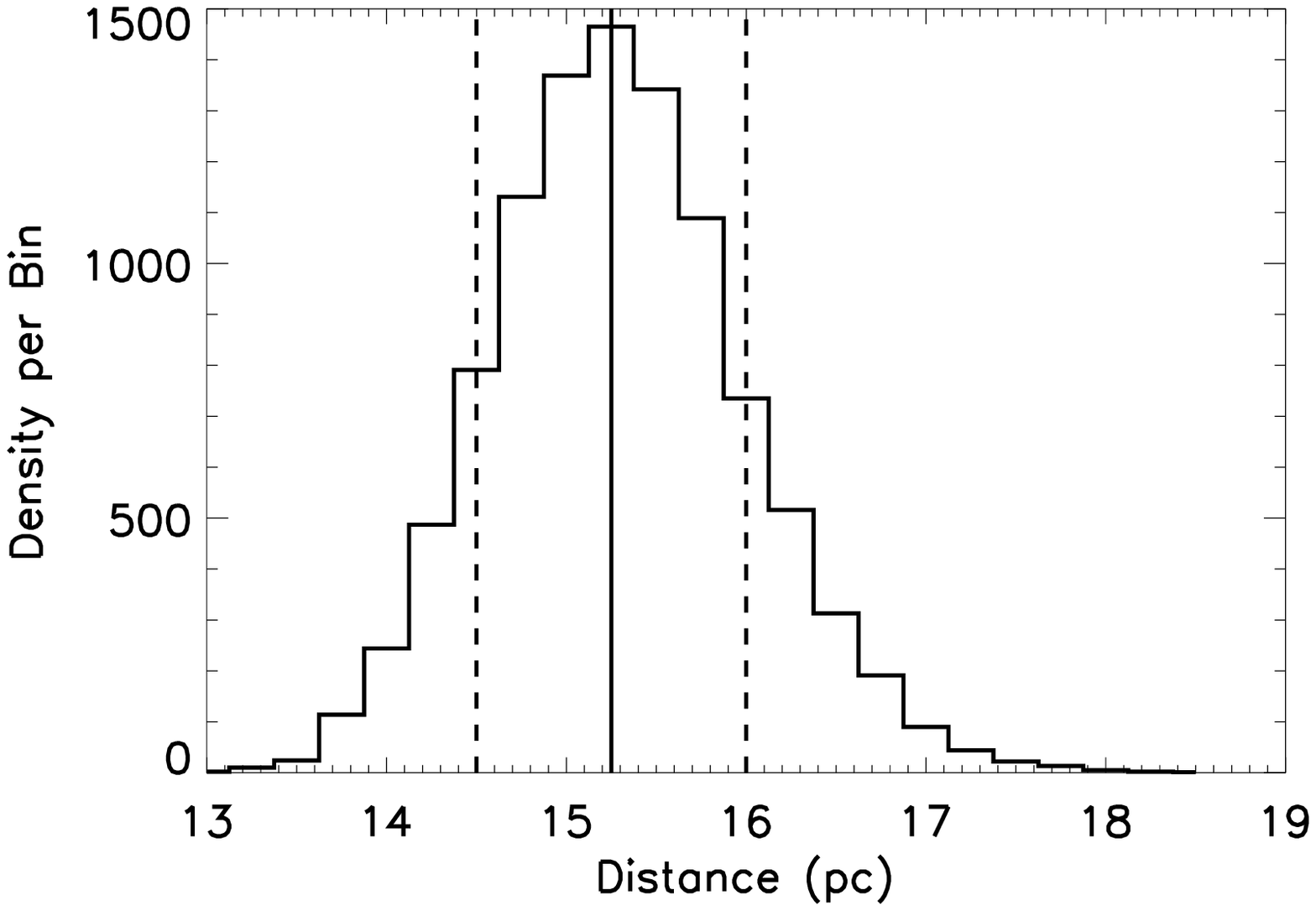} &
\includegraphics[width=3.in]{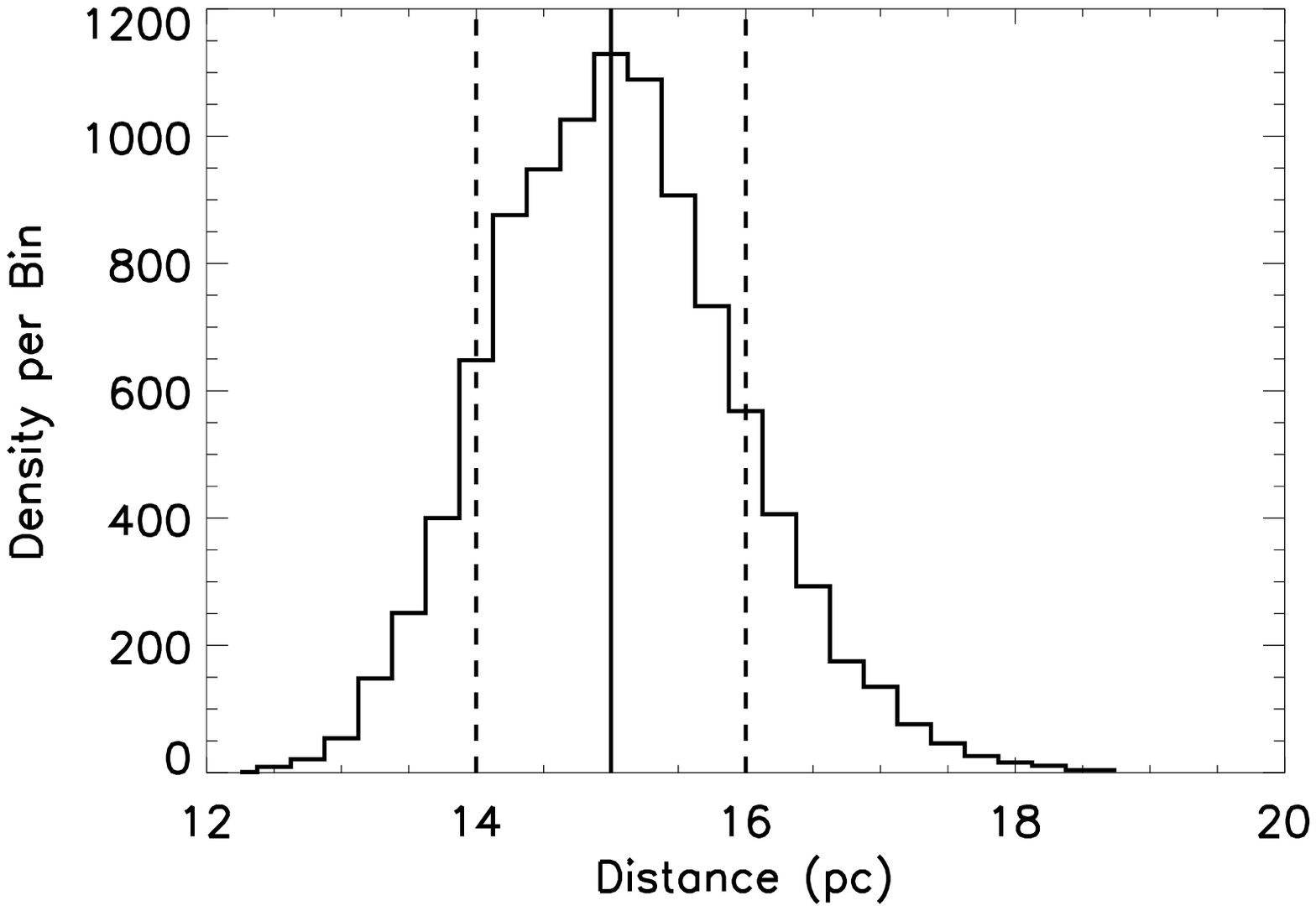} \\
\end{tabular}
\end{center}
\caption{Histograms of the distributions of Monte-Carlo simulation 
observation fits before absolute parallax correction.  
Upper left: fits to the RA component of 2M1207.
Upper right: fits to the DEC component of 2M1207.  Lower left: fits to the 
RA component of LHS 2397a.  
Lower right: fits to the DEC component of LHS 2397a.
The solid line marks the mean of each distribution and the dashed lines mark 
1$\sigma$ from the mean.}
\label{fig:hist}
\end{figure}

\begin{figure}
\begin{center}
\begin{tabular}{cc}
 \includegraphics[angle=0,width=3in]{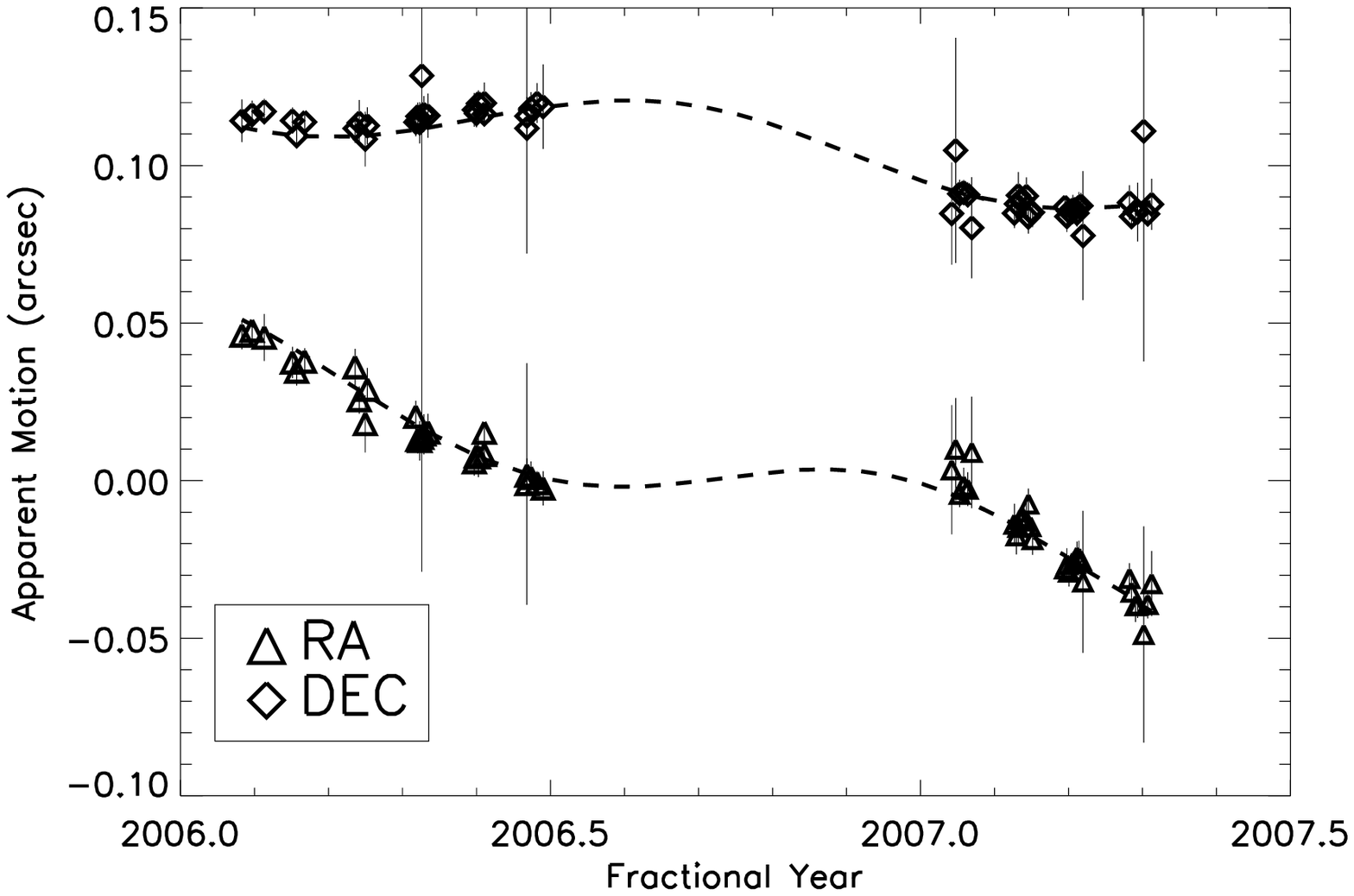} &
\includegraphics[angle=0,width=3in]{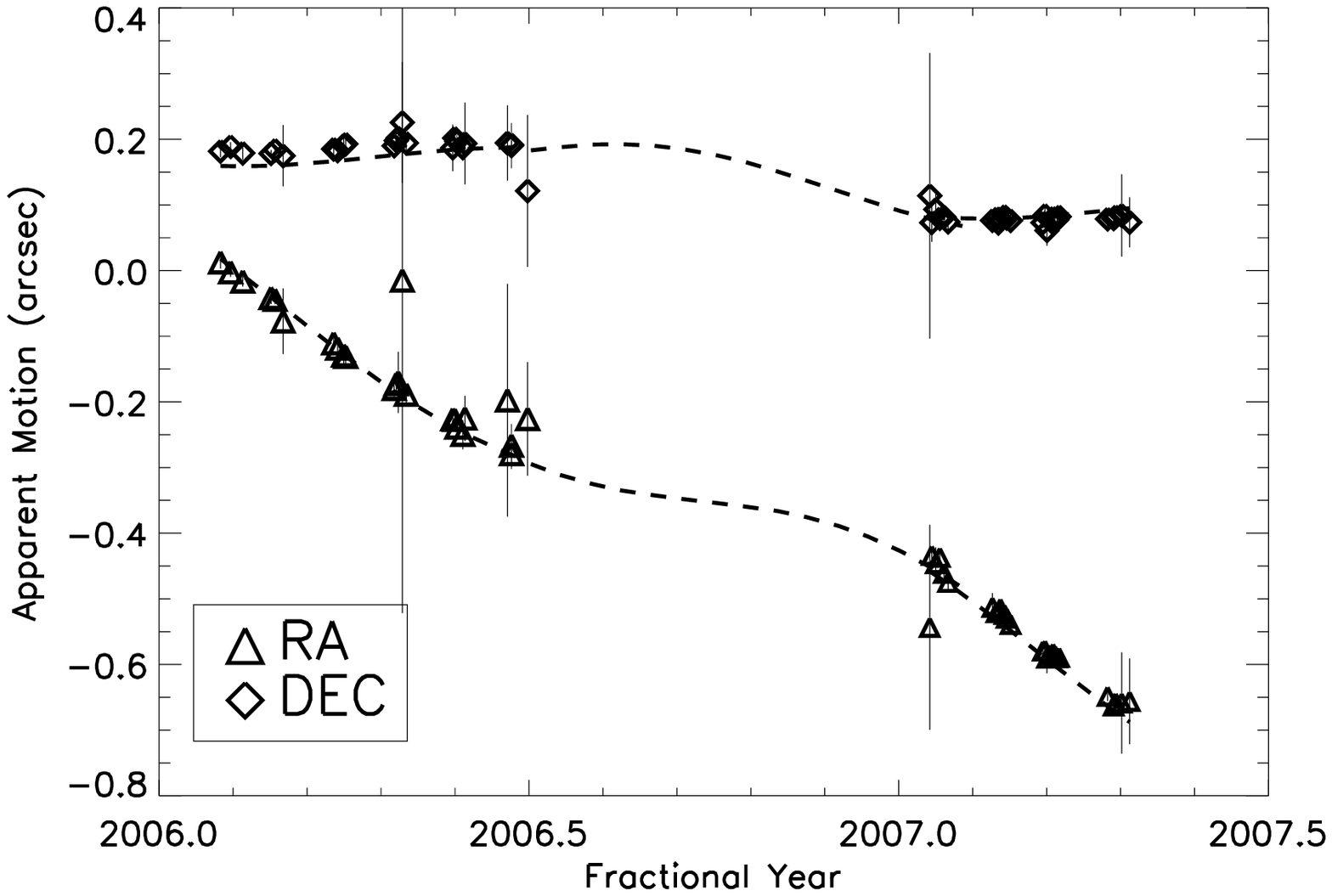} \\
\end{tabular}
\end{center}
\caption{RA and DEC measured relative motion curves (before absolute
parallax correction) for 2M1207A (left) and 
LHS 2397a (right).  The dashed
lines represent the 15.8 mas / 63 pc 
relative parallax best fit model for 2M1207A and 
the 66.7 mas / 15 pc relative parallax best fit model for LHS 2397a.
Correcting for absolute parallax, we find an 
absolute parallax of 17.0$^{+2.3}_{-1.8}$ mas and a 
best fit distance of 58.8$\pm$7.0 pc (all 1.28$\sigma$ errors) for 
2M1207A and an absolute parallax of 
67.9$^{+5.3}_{-4.7}$ mas (similar to the previous result of 62.6$\pm$4.0 mas, 
Tinney 1996) and a best fit distance of 14.7$\pm$1.0 pc for our
standard LHS 2397a.}
\label{fig:parallax}
\end{figure}

\end{document}